\documentclass[aps,prl,twocolumn,groupedaddress]{revtex4}
\usepackage{graphicx}
\usepackage{color}

\def\beq{\begin{equation}}
\def\eeq{\end{equation}}
\def\beqr{\begin{eqnarray}}
\def\eeqr{\end{eqnarray}}
\def\bdpm{\begin{displaymath}}
\def\edpm{\end{displaymath}}

\newcommand{\lsim}{\raisebox{-0.13cm}{~\shortstack{$<$ \\[-0.07cm] $\sim$}}~}

\begin{document}


\title{Longitudinal top polarization as a probe of 
a possible origin of forward-backward asymmetry 
of the top quark at the Tevatron}



\author{Dong-Won Jung}
\affiliation{Department of Physics, National Tsing Hua University
}
\affiliation{Physics Division, National Center for Theoretical Sciences, 
Hsinchu, Taiwan 300}

\author{P. Ko}
\affiliation{School of Physics, KIAS, Seoul 130-722, Korea}

\author{Jae Sik Lee}
\affiliation{Physics Division, National Center for Theoretical Sciences, 
Hsinchu, Taiwan 300}


\date{November 28, 2010}

\begin{abstract}
If the forward-backward (FB) asymmetry of top quark 
($A_{\rm FB}$) observed at the Tevatron deviates from the 
SM prediction, there must be $P$-violating 
interactions in $q\bar{q} \rightarrow t\bar{t}$. 
This new interaction will necessarily affect the top spin 
polarization.  In this letter,  
we perform a model independent analysis on the longitudinal 
(anti)top polarization ($P_L$ and $\bar{P}_L$) 
using an effective lagrangian with dim-6 four-quark 
operators relevant for $q \bar{q} \rightarrow t \bar{t}$, 
and show that the $P$-odd observable corresponding to
the polarization difference $(P_L - \bar{P}_L)$ 
gives important informations on the chiral structures of 
new physics that might be relevant to the $A_{\rm FB}$. 
\end{abstract}


\maketitle



1. Top physics has entered a new era after its first discovery, 
due to the high luminosity achieved at the Tevatron and 
the launch of the Large Hadron Collider (LHC).  
Most recent results on the top mass and the $t\bar{t}$ production 
cross section (CDF and D0 Collaborations combined analysis) are 
:
$m_t  =   ( 171.3 \pm 1.3 ) ~{\rm GeV}$  and 
$\sigma_{t\bar{t}}  =   ( 7.50 \pm 0.48 ) ~{\rm pb}$, respectively  
\cite{cdf2009}.
Being the heaviest particle observed so far with its mass being near
the electroweak breaking (EWSB) scale, the top sector might 
provide a new window to the EWSB mechanism.  
Precise determination of top quark properties is essential 
to address this issue, such as the top compositeness. 

The forward-backward asymmetry $A_{\rm FB}$ of the top quark is 
one of the interesting observables related with top quark. 
Within the SM, this asymmetry vanishes at leading order 
in  QCD because of $C$ symmetry. 
At next-to-leading order [$O(\alpha_s^3)$], 
a nonzero $A_{\rm FB}$  can develop from the interference
between the Born amplitude and two-gluon intermediate state, 
as well as the gluon bremsstrahlung and gluon-(anti)quark scattering 
into $t \bar{t}$, 
with the prediction $A_{\rm FB}\sim 0.078$ \cite{Antunano:2007da}.  
The measured asymmetry has been off the SM prediction by $2 \sigma$ 
for the last few years, albeit a large experimental uncertainties. 
The most recent measurement in the $t\bar{t}$ rest frame is 
\cite{cdf2010}
\begin{eqnarray}
A_{\rm FB} & \equiv & \frac{N_t ( \cos\theta \geq 0) - N_{\bar{t}} 
( \cos\theta \geq 0 )}{N_t ( \cos\theta \geq 0) + N_{\bar{t}} 
( \cos\theta \geq 0 )} \\
& = & 
(0.158 \pm 0.072  \pm 0.017) 
\end{eqnarray}
with $\theta$ being the polar angle of the top quark with 
respect to the incoming proton in the $t\bar{t}$ rest frame.
The newest number is somewhat lower than the previous one
\cite{cdf2009},  $A_{\rm FB} = 0.24 \pm 0.13 \pm 0.04$, 
which  had stimulated a lot of activities on possible new physics scenarios 
\cite{Choudhury:2007ux,Djouadi:2009nb,Ferrario:2009bz,Jung:2009jz,
Cheung:2009ch,Frampton:2009rk,Shu:2009xf,Arhrib:2009hu,Ferrario:2009ee,Dorsner:2009mq,jkln1,
Cao:2009uz,Barger:2010mw,Cao:2010zb,Xiao:2010hm,Martynov:2010ed,
Chivukula:2010fk,Rodrigo:2010gm,Bauer:2010iq,Chen:2010hm,Xiao:2010ph,Wang:2010tg} .

Since the central value of the $A_{\rm FB}$ is getting closer 
to the SM prediction, any new physics effects might be smaller 
than had been thought previously.
Also  there is no clear signal for such a new resonance \cite{cdf2009}.  
Therefore, it would be reasonable to assume a new physics scale relevant 
to $A_{\rm FB}$ is large enough so that  production of a new particle is 
beyond the reach of the Tevatron \cite{jkln1}, 
which makes a key difference between our work and other literatures. 
Then it is adequate to integrate out the heavy fields, and we can adopt 
a model independent effective lagrangian approach in order to study 
new physics effects on $\sigma_{t\bar{t}}$ and $A_{\rm FB}$.  
If new physics scale is high enough, then their effects on the $t\bar{t}$ 
production at the Tevatron can be described by dim-6 effective lagrangian. 
Since the $SU(3)_C \times SU(2)_L \times U(1)_Y $
symmetry has been well established for the light quark system,  
we assume that $SU(2)_L \times U(1)_Y$ symmetry is linearly realized
on the light quark system.  And we impose the custodial symmetry 
$SU(2)_R$ for the light quark sector.  
Under these assumptions, the dimension-6 operators relevant to the 
$t\bar{t}$ production at the Tevatron are
\begin{eqnarray}
\mathcal{L}_6 &=& \frac{g_s^2}{\Lambda^2}\sum_{A,B}
\left[C^{AB}_{1q}(\bar{q}_A\gamma_\mu  q_A)
(\bar{t}_B\gamma^\mu t_B) \right.
\nonumber \\
&& \hspace{1.0cm} + \left.
C^{AB}_{8q}(\bar{q}_A T^a\gamma_\mu q_A)(\bar{t}_B 
T^a\gamma^\mu  t_B)\right]
\label{eq:L_6eff}
\end{eqnarray}
where $T^a = \lambda^a /2$, $\{A,B\}=\{L,R\}$, and 
$L,R \equiv (1 \mp \gamma_5)/2$ 
with $q=(u,d)^T,(s,c)^T$ 
\footnote{Although we assume the $SU(2)_L \times SU(2)_R$ chiral symmetry
for light quarks, all the explicit models do not satisfy this condition. 
In that case, one can interpret $q=u,d,s,c,b$.}.
Our choice of dim-6 operators is basically the same as 
Ref.~\cite{Hill:1993hs}, except that we use the chiral basis for $t$ 
and $\bar{t}$.  This operator set could be used, for example, to study
$t\bar{t}$ production at the Tevatron in case of the composite top 
scenarios \cite{Georgi:1994ha}. 

Before we move to the main subject of this paper, we would like to 
make a comment on other dim-6 operators that involves $t$ and $\bar{t}$.
In principle, there are many more operators that involve $t$, $\bar{t}$ and 
gluon field strength tensor $G_{\mu\nu}^a$, 
which have been studied  recently in Refs~\cite{willenbrock} and 
\cite{AguilarSaavedra:2010zi}. 
Many of them are however generated at one-loop level, unlike the operators 
we are considering here and in Ref.~\cite{jkln1}.  
Therefore their effects would be further suppressed by a loop factor 
$1/(4\pi)^2$ and a power of strong coupling constant $g_s$, 
relative to the operators we study. 
Our choice of operators should be enough for the purpose of 
$t\bar{t}$ production at the Tevatron. 

Using the above effective lagrangian, we can calculate the cross section 
up to $O(1/\Lambda^2)$, keeping only the interference term between 
the standard model and new physics contributions.  
The squared amplitude summed (averaged) over the final (initial) 
spins and colors is given by
\begin{widetext}
\begin{eqnarray}
\overline{|{\cal M}|^2_0} 
\simeq  \frac{4\,g_s^4}{9\,\hat{s}^2} \left\{
2 m_t^2 \hat{s} \left[
1+\frac{\hat{s}}{2\Lambda^2}\,(C_1+C_2)
\right] s_{\hat\theta}^2 
+\frac{\hat{s}^2}{2}\left[ \left(1+\frac{\hat{s}}{2\Lambda^2}\,(C_1+C_2)\right)
(1+c_{\hat\theta}^2)
+\hat\beta_t\left(\frac{\hat{s}}{\Lambda^2}\,(C_1-C_2)\right)c_{\hat\theta}
\right]\right\}
\label{eq:amp0sq}
\end{eqnarray}
\end{widetext}
where $\hat{s} = (p_1 + p_2)^2$, $\hat\beta_t^2=1-4m_t^2/\hat{s}$,
and $s_{\hat\theta}\equiv \sin\hat\theta$ and 
$c_{\hat\theta}\equiv \cos\hat\theta$ 
with $\hat{\theta}$ being the polar
angle between the incoming quark and the outgoing top quark in the 
$t\bar{t}$ rest frame. 
And two couplings $C_{1,2}$ are defined as
\footnote{Throughout this work, unless explicitly written, we are taking
$C^{AB}_{8q}=C^{AB}_{8u}=C^{AB}_{8d}$ assuming the 
$SU(2)_L \times SU(2)_R$ chiral symmetry. Under this assumption, 
the down-quark contribution to $\sigma_{t\bar{t}}$ and
$A_{\rm FB}$ is suppressed relative to the 
up-quark one by a factor more than $\sim 6$ at the Tevatron.}
\begin{equation}
C_1 \equiv C_{8q}^{RR}+C_{8q}^{LL}\,, \ \ \
C_2 \equiv C_{8q}^{LR}+C_{8q}^{RL} \,.
\end{equation}

\begin{figure}
\hspace{-1.5cm}
\includegraphics[width=10cm]{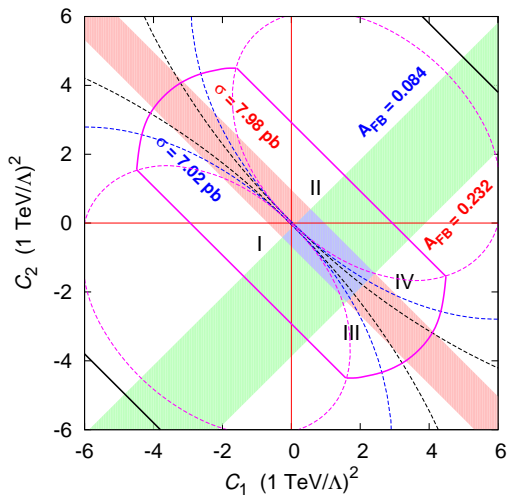}
\caption{
The region in $( C_1 , C_2 )$ plane that is
consistent with the Tevatron data  at the 1-$\sigma$ level:
$\sigma_{t\bar{t}} = (7.50 \pm 0.48)$ pb
and $A_{\rm FB} = (0.158 \pm 0.072 \pm 0.017)$.
Also shown are the boundaries of the regions 
where our effective lagrangian description is valid.
For details, we refer to Ref.~\cite{jkln1}.
}
\label{fig:newfig1}
\end{figure}

In our previous study \cite{jkln1}, we performed a model independent
study of $\sigma_{t\bar{t}}$ and $A_{\rm FB}$ considering the interference
effects of the SM amplitude and the new physics amplitudes from 
dim-6 operators, the leading order operators in the effective lagrangian.
Here we update the previous results in the light of the new measurement
of $A_{\rm FB}$, see Fig.~\ref{fig:newfig1}.
The main results of Ref.~\cite{jkln1} can be summarized as follows
in terms of two effective couplings $C_1$ and $C_2$:
%
\begin{itemize}
\item $\Delta\sigma_{t\bar{t}}\equiv
\sigma_{t\bar{t}} - \sigma_{t\bar{t}}^{\rm SM} \propto ( C_1 + C_2 )$, 
whereas $\Delta A_{\rm FB}\equiv 
A_{\rm FB}-A_{\rm FB}^{\rm SM} \propto (C_1 - C_2)$, 
i.e., the new physics contributions to 
the total cross section and $A_{\rm FB}$ are orthogonal. 
Therefore the new physics can change $A_{\rm FB}$ considerably without
affecting $\sigma_{t\bar{t}}$ too much, as long as $C_1 + C_2 \approx 0$. 
\item In order to have nonzero new physics contribution to $A_{\rm FB}$, 
we need  $C_1 - C_2 \neq 0$.
If parity were conserved in the light quark sector in dim-6 operators, 
one would have $C^{LL}_{8q} = C^{RL}_{8q}$, and $C^{LR}_{8q} = C^{RR}_{8q}$. 
If parity were conserved in the top quark sector, one would have 
$C^{LL}_{8q} = C^{LR}_{8q}$ and $C^{RR}_{8q} = C^{RL}_{8q}$. 
In either case,  we end up with
the vanishing condition: $(C_1 - C_2) = 0$. 
Therefore, in order to nonzero new physics contribution from dim-6 operators, 
one has to break parity $P$ both in the light quark and the top quark sectors. 
This might be observable in (or constrained by) parity violating effects 
in nucleon nuclear scattering, for example.

\item The usual spin-spin correlation $C$ is strongly correlated with the 
top quark pair production cross section $\sigma_{t\bar{t}}$, and not with
the $A_{\rm FB}$. On the other hand, the newly defined FB spin-spin correlation
$C_{\rm FB}$ is strongly correlated with the $A_{\rm FB}$, and thus can be 
another important check of any anomaly in $A_{\rm FB}$. If there is any deviation
in $A_{\rm FB}$, should there be some deviation in $C_{\rm FB}$ too.
\item Since $\sigma_{t\bar{t}}$ and $A_{\rm FB}$ depend only on two combinations
$C_1$ and $C_2$,  we can not know exactly the chiral structure of new physics 
from these two observables alone.
We need another physical observables which are sensitive to independent 
combinations of coupling constants in dim-6 operators. 
\end{itemize}
It is the purpose of this letter to present new observables which show different
dependence on $C^{AB}_{8q}$'s 
from $\sigma_{t\bar{t}}$ and $A_{\rm FB}$. 
What we propose is the longitudinal polarization of top quark, 
$P_L \equiv \langle \vec{S_t} \cdot \vec{n}_t
\rangle$, where $\vec{n}_t$ is 
any unit vector defining the spin quantization axis of the top quark, 
and similarly for the antitop:  
$\bar{P}_L \equiv \langle \vec{S_{\bar{t}}} \cdot 
\vec{n}_{\bar{t}} \rangle$.
If we choose $\vec{n}_{t\,(\bar{t})} = 
\vec{p}_{t\,(\bar{t})} / | \vec{p}_{t\,(\bar{t})} |
$ with $\vec{p}_{t\,(\bar{t})}$ being the
momentum vector of $t$ ($\bar{t}$),
$P_L$ ($\bar{P}_L$) becomes the usual helicity of (anti)top quark.  
Any observables corresponding to the 
longitudinal-polarization combinations
$(P_L\pm \bar{P}_L)$ vanish
in QCD because of parity ($P$) conservation. 
On the other hand, if there is new physics that affects $A_{\rm FB}$,
parity is necessarily broken. Therefore one can expect nonzero 
$P$-violating polarization observables in general,
which is the main point of the present work. 

\bigskip


2. Now let us study the polarizations of $t$ and $\bar{t}$ at the 
Tevatron using the helicity amplitude method. In particular, 
we consider the polarization coefficients involving the
longitudinal polarizations of $t$ and $\bar{t}$
which vanish in QCD due to its $P$ conservation. 

In the center-of-mass frame of the $t\bar{t}$ pair,
the helicity amplitudes for the process 
$q(\lambda)\bar{q}(\bar\lambda) \rightarrow t(\sigma)\bar{t}(\bar\sigma)$
induced by the dimension-6 operators, Eq.~(\ref{eq:L_6eff}), 
and as well as the SM interactions are given by
\begin{eqnarray}
{\cal M}(\sigma,\bar\sigma;\lambda,\bar\lambda) &\equiv &
\frac{g_s^2}{\hat{s}} \left[
\delta_{ij}\delta_{kl}\,
\langle\sigma,\bar\sigma;\lambda,\bar\lambda\rangle _{\rm sing}  \right.
\nonumber \\
&& \hspace{0.4cm} \left. + T^a_{ij}T^a_{kl}\,
\langle\sigma,\bar\sigma;\lambda,\bar\lambda\rangle _{\rm oct}\right]
\end{eqnarray}
where we denote the helicities of the incoming quarks by $\lambda$ and 
$\bar\lambda$ and those of the outgoing top quarks by
$\sigma$ and $\bar\sigma$, respectively, with 
$\lambda\,,\sigma=+$ and $-$ standing for right- and left-handed particles.
The singlet amplitude 
$\langle\sigma,\bar\sigma;\lambda,\bar\lambda\rangle _{\rm sing}$
is irrelevant in our case where we keep only the interference term between
the SM and the new physics contributions.
The octet amplitude 
$\langle\sigma,\bar\sigma;\lambda,\bar\lambda\rangle _{\rm oct}$ 
can be written as
\begin{eqnarray}
\langle\sigma,\bar\sigma;\lambda,\bar\lambda\rangle _{\rm oct} \equiv 
\sum_{A,B=L,R}\,
\left(1+\frac{\hat{s}}{\Lambda^2}\,C^{AB}_{8q}\right)\,
\langle\sigma,\bar\sigma;\lambda,\bar\lambda\rangle _{AB}^V
\label{eq:octet_amp}
\end{eqnarray}
where the first and the second terms count for 
the contributions from the SM QCD and the dim-6 operators,
respectively.
The reduced amplitudes 
$\langle\sigma,\bar\sigma;\lambda,\bar\lambda\rangle _{AB}^V$
are explicitly given by 
\begin{eqnarray}
\langle\sigma,\bar\sigma;\lambda,\bar\lambda\rangle _{AB}^V
&\equiv &
-\frac{m_t\sqrt{\hat{s}}}{2}\,(1+A\lambda)\,\sigma\, s_{\hat\theta}\,
\delta_{\lambda,-\bar\lambda}\,\delta_{\sigma,\bar\sigma}
\nonumber \\
&&
- \frac{\hat{s}}{4}\Bigg[
(1+A\lambda)(1+\hat\beta_t B \sigma)\,c_{\hat\theta}
\\ && 
\hspace{0.5cm} + (A+\lambda)(\hat\beta_t B+\sigma) \Bigg]
\delta_{\lambda,-\bar\lambda}\,\delta_{\sigma,-\bar\sigma}\,. \nonumber
\end{eqnarray}
The top-polarization weighted squared matrix elements can be
computed from the helicity amplitudes by a suitable
rotation~\cite{Hagiwara:1985yu} from the helicity basis to a general spin basis:
\begin{eqnarray}
\overline{|{\mathcal{M}}|^2} = 
\frac{2}{9}\,\frac{g_s^4}{\hat{s}^2}\,\sum_{\lambda,\bar\lambda}
\left\{
{\rm Tr}[\langle\sigma,\bar\sigma;\lambda,\bar\lambda\rangle _{\rm oct}\, \bar{\rho}^T\,
\langle\sigma,\bar\sigma;\lambda,\bar\lambda\rangle _{\rm oct}^\dagger\,\rho]
\right\}
\label{eq:trace}
\end{eqnarray}
where $\rho$ and $\bar{\rho}$ are $2\times 2 $ polarization density matrices
for the top and anti-top, respectively:
\begin{eqnarray}
\rho &=& \frac{1}{2}\left(\begin{array}{cc} 1+P_L  & P_T e^{-i\alpha}\\ P_T e^{i\alpha}& 1-
P_L \end{array}\right), \nonumber \\
\bar{\rho} &=& \frac{1}{2}\left(\begin{array}{cc}
1+\bar{P}_L  & -\bar{P}_T e^{i\bar{\alpha}}\\ -\bar{P}_T e^{-i\bar{\alpha}}& 1- \bar{P}_L
\end{array}\right).
\end{eqnarray}
Here, $P_L$ and $\bar{P}_L$ are the longitudinal polarizations of $t$ and $\bar{t}$,
respectively,
while $P_T$ and $\bar{P}_T$ the degrees of transverse polarization with
$\alpha$ and $\bar\alpha$ being the azimuthal angles
with  respect to the $t$-$\bar{t}$  production plane.

Neglecting the transverse polarizations, 
an expansion of the trace in Eq.~(\ref{eq:trace}) leads to
\begin{eqnarray}
\overline{|{\mathcal{M}}|^2} & = & \frac{g_s^4}{\hat{s}^2}\Bigg\{
{\cal D}_0  +
{\cal D}_1 (P_L+\bar{P}_L) 
\nonumber \\ && \hspace{0.5cm} + 
{\cal D}_2 (P_L-\bar{P}_L)+
{\cal D}_3 P_L \bar{P}_L \Bigg\}\,.
\end{eqnarray}
The polarization coefficients ${\cal D}_i (i = 0 - 3)$ are defined in terms of the 
octet helicity amplitudes by
\begin{eqnarray}
{\cal D}_ 0 &=& \frac{2}{9}\cdot\frac{1}{4}\sum_{\lambda\,,\bar\lambda }\bigg (|\langle ++;
\lambda\bar\lambda  \rangle_{\rm oct}|^2 
+ |\langle --; \lambda\bar\lambda  \rangle_{\rm oct}|^2 
\nonumber \\ && \hspace{1.4cm}
+ |\langle +-; \lambda\bar\lambda
\rangle_{\rm oct}|^2 + |\langle -+;
\lambda\bar\lambda  \rangle_{\rm oct}|^2 \bigg ),\nonumber\\
{\cal D}_ 1 &=& \frac{2}{9}\cdot\frac{1}{4}\sum_{\lambda\,,\bar\lambda }\bigg (|\langle ++;
\lambda\bar\lambda  \rangle_{\rm oct}|^2 -
|\langle --;
\lambda\bar\lambda  \rangle_{\rm oct}|^2 \bigg)\,, \nonumber\\
{\cal D}_ 2 &=& \frac{2}{9}\cdot\frac{1}{4}\sum_{\lambda\,,\bar\lambda }\bigg (
|\langle +-; \lambda\bar\lambda  \rangle_{\rm oct}|^2 - |\langle -+; \lambda\bar\lambda
\rangle_{\rm oct}|^2
\bigg ),\nonumber\\
{\cal D}_ 3 &=& \frac{2}{9}\cdot\frac{1}{4}\sum_{\lambda\,,\bar\lambda }\bigg (|\langle ++;
\lambda\bar\lambda  \rangle_{\rm oct}|^2 +
|\langle --; \lambda\bar\lambda  \rangle_{\rm oct}|^2 
\nonumber \\ && \hspace{1.4cm}
- |\langle +-; \lambda\bar\lambda
\rangle_{\rm oct}|^2 - |\langle -+;
\lambda\bar\lambda  \rangle_{\rm oct}|^2 \bigg ),
\end{eqnarray}
The unpolarized coefficient ${\cal D}_0$ gives
the squared amplitude summed
(averaged) over the final (initial) spins and colors and 
one may obtain the same expression as Eq.~(\ref{eq:amp0sq}) 
by keeping
the terms up to $O(1/\Lambda^2)$ in ${\cal D}_0$.
So, the unpolarized coefficient ${\cal D}_0$ leads to
the total cross section $\sigma_{t\bar{t}}$ and the forward-backward
asymmetry $A_{\rm FB}$.
On the other hand, the coefficient ${\cal D}_3$ gives the 
spin-spin correlations $C$ and $C_{\rm FB}$ considered 
and suggested before.

Note that
the other two coefficients ${\cal D}_1$ and ${\cal D}_2$
are $P$ violating. Furthermore, the coefficient ${\cal D}_1$ 
is odd  under both the CP and CP$\widetilde{\rm T}$ transformations
\footnote{The $\widetilde{\rm T}$ transformation
reverses the signs of the spins
and the three-momenta of the
asymptotic states, without interchanging initial and final
states, and the matrix element gets complex conjugated.}.
In our effective lagrangian approach, new heavy particles 
are integrated out, and there is no new strong CP-even phase, and so 
${\cal D}_1$ is zero. However, it could be nonzero when 
the heavy particle is explicitly included, and we keep
the finite decay width of the heavy particle 
together with possible CP-violating phases in its couplings to
light and top quarks.   This issue will be discussed in full 
in the future publication \cite{progress}.

\begin{figure}
\hspace{-1.5cm}
\includegraphics[width=10cm]{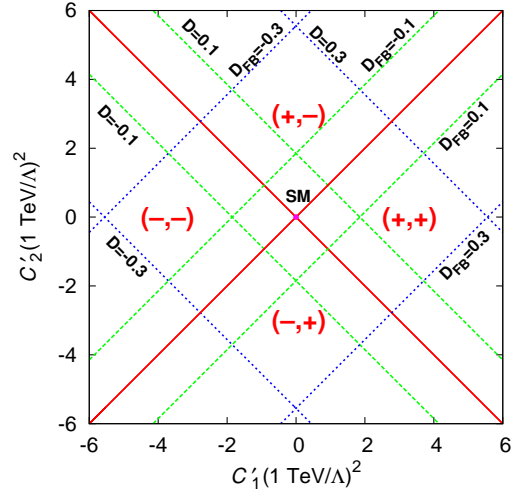}
\caption{
The $P$-violating spin correlations $D$ and
$D_{\rm FB}$ in the $(C_1^\prime,C_2^\prime)$ plane.
The signs of $(D\,,D_{\rm FB})$ are denoted.}
\label{fig:ddfb}
\end{figure}

The other $P$-violating coefficient
${\cal D}_2$ could be observable at the Tevatron, revealing
genuine features of new physics responsible for $A_{\rm FB}$.
Explicitly, we have obtained 
\begin{equation}
{\cal D}_2 \simeq \frac{\hat{s}}{9\,\Lambda^2}\left[
(C_1^\prime + C_2^\prime) \hat{\beta_t} (1+c^2_{\hat{\theta}})
 +(C_1^\prime - C_2^\prime) (5-3 \hat{\beta}_t^2) c_{\hat{\theta}}\right]
\end{equation}
with
\begin{eqnarray}
C_1^\prime  \equiv  C_{8q}^{RR}-C_{8q}^{LL}\,, \quad
C_2^\prime  \equiv  C_{8q}^{LR}-C_{8q}^{RL}\,.
\end{eqnarray}
Therefore ${\cal D}_2$ will provide additional information
on the chiral structure of new physics in $q\bar{q} \rightarrow t\bar{t}$.
When we integrate over the polar angle $\hat{\theta}$, only the first term 
involving 
\[
(C_1^\prime + C_2^\prime) = C_{8q}^{RR} - C_{8q}^{LL} + C_{8q}^{LR} - C_{8q}^{RL} 
\]
survives. 
On the other hand, if we separate the forward and the backward top 
samples and take the difference, the orthogonal combination in the second term 
survives:
\[
(C_1^\prime - C_2^\prime) = C_{8q}^{RR} - C_{8q}^{LL} - C_{8q}^{LR} + C_{8q}^{RL} \,.
\]
For definiteness, we consider the two new observables:
\begin{eqnarray}
D &\equiv & \frac{\sigma(t_R\bar{t}_L) - \sigma(t_L\bar{t}_R)}
{\sigma(t_R\bar{t}_R) + \sigma(t_L\bar{t}_L) +
\sigma(t_L\bar{t}_R) + \sigma(t_R\bar{t}_L)}\,,
\nonumber \\[0.1cm]
D_{\rm FB} &\equiv &
D (\cos\hat\theta \geq 0) -  D (\cos\hat\theta \leq 0)
\end{eqnarray}
which involve the sum and difference of the coefficients
$C_1^\prime$ and $C_2^\prime$, respectively.
In Fig.~\ref{fig:ddfb}, we show the $P$-violating spin correlations
$D$ and $D_{\rm FB}$ in in the $(C_1^\prime,C_2^\prime)$ plane. 
We observe that $|D|$ and $|D_{\rm FB}|$ could be as large as $0.1$
in the region $|C_{1,2}^\prime\,(1\,{\rm TeV}/\Lambda)^2|\lsim 1$ which
can be observed with an event sample of about 100 $t\bar{t}$ pairs
after event selection cuts.
Note that there are no experimental constraints on the
$D$ and $D_{\rm FB}$ observables yet,
but they can be measured with a statistical precision of $\sim 5$ \%
using the full anticipated Tevatron data set of 
10 fb$^{-1}$~\cite{Amidei}.

In principle, the polarization coefficients could be measured by
studying the angular distributions of the
top-quark decay products. The top and anti-top quarks 
decay into two $b$ quarks and two $W$ bosons. When 
both of the $W$ bosons decay leptonically, 
in the helicity basis, the amplitude squared can be written as
\begin{eqnarray}
\overline{|{\mathcal{M}}|^2} & = & \frac{g_s^4}{\hat{s}^2}\Bigg\{
{\cal D}_0  +
{\cal D}_1 (\cos\theta^*_+ + \cos\theta^*_-) 
\nonumber \\ && \hspace{0.0cm} + 
{\cal D}_2 (\cos\theta^*_+ - \cos\theta^*_-)+
{\cal D}_3 \cos\theta^*_+ \cos\theta^*_- \Bigg\}\,.
\end{eqnarray}
where $\theta^*_+$ ($\theta^*_-$)
is the angle between the charged lepton $l^+$ $(l^-)$
in the top (anti-top) rest frame and
the direction of the top (anti-top) in the $t\bar{t}$ rest frame.
The $M_{T2}$ variable could be useful to reconstruct
the $t\bar{t}$ rest frame even with the two missing neutrinos,
which deserves a further study in the future.

\bigskip 


3. Now we study specific new physics that could generate the 
relevant dim-6 operators with corresponding Wilson coefficients.
It is impossible to exhaust all the possibilities, and we consider
the following interactions of quarks with spin-1 
flavor-conserving (FC) color-octet $V^a_{8A}$ vectors,
spin-1 flavor-violating (FV) color-singlet $\tilde{V}_{1A}$ and color-octet 
$\tilde{V}^a_{8A}$ vectors, and spin-0 FV color-singlet $\tilde{S_1}$
and color-octet $\tilde{S}^a_{8A}$ scalars ($A=L,R$):
\beqr
&& \mathcal{L}_{\rm int} =  
g_s \sum_{A} V_{8A}^{a\mu}\left[g_{8q}^A(\bar{q}_A\gamma_\mu T^a q_A) 
+ g_{8t}^A(\bar{t}_A\gamma_\mu T^a t_A) \right]    
\nonumber  \\
&& 
+ g_s\sum_{A} \big[\tilde{V}_{1A}^\mu \tilde{g}_{1q}^A(\bar{t}_A\gamma_\mu q_A)
+ \tilde{V}_{8A}^{a\mu}\tilde{g}_{8q}^A(\bar{t}_A\gamma_\mu T^a q_A) 
+ \textrm{h.c.} \big]    \nonumber \\
&& + g_s \sum_{A}\big[\tilde{S}_{1A} \tilde{\eta}_{1q}^A(\bar{t} A q)
+ \tilde{S}_{8A}^{a} \tilde{\eta}_{8q}^A(\bar{t} A T^a q) + \textrm{h.c.} \big] ,
\eeqr 
%
where $q$ denotes light quarks (either $u$ or $d$ depending on the models).
This interaction lagrangian encompasses many models beyond the SM, and 
make a good starting point to study the underlying mechanism for 
the effective lagrangian discussed earlier.
If the spin-1 particle has both the FC and FV interactions,  
we may set $V_{8}^\mu =\tilde{V}_{8}^\mu$.

After integrating out the heavy vector and scalar fields, we obtain 
the Wilson coefficients as follows:
\beqr
\frac{C^{RR}_{8q}}{\Lambda^2} &=& -\frac{g_{8q}^R g_{8t}^{R}}{m_{V_{8R}}^2} 
 - \frac{2 |\tilde{g}_{1q}^R |^2}{m^2_{\tilde{V}_{1R}}}
 + \frac{1}{N_C}\,\frac{|\tilde{g}_{8q}^R |^2}{m^2_{\tilde{V}_{8R}}}\,,  
\label{eq:wcs1} \\
\frac{C^{LL}_{8q}}{\Lambda^2} &=& -\frac{g_{8q}^L g_{8t}^{L}}{m_{V_{8L}}^2} 
 - \frac{2 |\tilde{g}_{1q}^L |^2}{m^2_{\tilde{V}_{1L}}}
 + \frac{1}{N_C}\,\frac{|\tilde{g}_{8q}^L |^2}{m^2_{\tilde{V}_{8L}}}\,,  
\nonumber   \\
\frac{C^{LR}_{8q}}{\Lambda^2} &=& -\frac{g_{8q}^L g_{8t}^{R}}{m_{V_8}^2} 
- \frac{|\tilde{\eta}_{1q}^L|^2}{m^2_{\tilde{S}_{1L}}} 
+ \frac{1}{2N_c}\frac{|\tilde{\eta}_{8q}^L|^2}{m^2_{\tilde{S}_{8L}}}\,, \nonumber  \\
\frac{C^{RL}_{8q}}{\Lambda^2} &=& -\frac{g_{8q}^R g_{8t}^{L}}{m_{V_8}^2} 
- \frac{|\tilde{\eta}_{1q}^R|^2}{m^2_{\tilde{S}_{1R}}} 
+ \frac{1}{2N_c}\frac{|\tilde{\eta}_{8q}^R|^2}{m^2_{\tilde{S}_{8R}}}\,, 
\nonumber  
\eeqr
where $m_{V_{8R,8L}}$ ($m_{\tilde{V}_{iR,iL}}$) and
$m_{\tilde{S}_{iR,iL}}$ denote the masses of vectors 
$V_{8R,8L}$ ($\tilde{V}_{iR,iL}$) and 
scalars $\tilde{S}_{iR,iL}$, respectively, with $i=1,8$.
Note that the contributions to the coefficients $C^{LR}_{8q}$ and $C^{RL}_{8q}$
from the FC color-octet vectors
may not be vanishing in the coexistence of $V_{8R}$ and $V_{8L}$  and in this case
we take $m_{V_{8R}}=m_{V_{8L}}=m_{V_{8}}$.

Another interesting possibility is minimal flavor violating interactions of 
color-triplet $S_k^\gamma$ with mass $m_{S_3}$ and 
color-sextet scalars $S_{ij}^{\alpha\beta}$ with mass
$m_{S_6}$ with with the SM quarks \cite{Arnold:2009ay}. Here 
$\alpha, \beta, \gamma$ and $i,j,k$ are color and flavor indices, respectively. 
For example, if we consider the following interactions (Model V and VI in 
Ref.~\cite{Arnold:2009ay}), 
\beq
{\cal L} = g_s \Big[ \frac{\eta_3}{2} \epsilon_{\alpha\beta\gamma} \epsilon^{ijk} 
u_{iR}^\alpha u_{jR}^\beta S_k^\gamma 
+  \eta_6 u_{iR}^\alpha u_{jR}^\beta S_{ij}^{\alpha\beta} + h.c. \Big]
\eeq 
the $u-$channel exchange of new scalars can contribute to 
$u \bar{u} \rightarrow t \bar{t}$, resulting in 
\footnote{ 
$C_{1q}^{RR}$ is also induced by color-triplet and 
sextet scalars,  but is not shown, since it is irrelevant here.
} 
\beq 
\label{eq:wcs2}
\frac{C_{8u}^{RR}}{\Lambda^2}   =  - \frac{| \eta_3 |^2}{ m_{S_3}^2} +  
\frac{2 |\eta_6|^2}{m_{S_6}^2} \ .
\eeq 
Since these new scalars couple only to the right-handed up-type quarks, 
constraints on the couplings $\eta_3$ and $\eta_8$ from flavor physics
are rather weak, and one can accommodate the observed $A_{\rm FB}$ easily.

In Table~\ref{tab:newparticles}, we show the new particle exchanges  
under consideration and the signs of the couplings induced by them.
Note that the particle exchanges with $(C_1-C_2)>0$ are preferred 
by the positive $A_{\rm FB}$ at the 1-$\sigma$ level.
%
\begin{table*}[t]
\caption{\label{tab:newparticles}
{\it
New particle exchanges and the signs of induced couplings
$C^{AB}$ ($A,B=R,L$),
$C_1-C_2$,
$C_1^\prime+C_2^\prime$, and
$C_1^\prime-C_2^\prime$.}
}
\begin{center}
\begin{tabular}{|c||c|c|c|c||c|c|c||c|}
\hline\hline
& & & & & & & &   \\[-0.3cm]
Resonance & $C^{RR}$ & $C^{LL}$ & $C^{LR}$ & $C^{RL}$ &
$C_1 - C_2$ &
$C_1^{'} + C_2^{'}$ & $C_1^{'} - C_2^{'}$ & $A_{\rm FB}$
\\[0.1cm]
\hline\hline
& & & & & & & & \\[-0.3cm]
$\tilde{V}_{1R}$  &  $-$ & 0 & 0 & 0 &
$-$ & $-$ & $-$ & $\times$  \\[0.1cm]
$\tilde{V}_{1L}$  &  0 & $-$ & 0 & 0 &
$-$ & $+$ & $+$ & $\times$  \\[0.1cm]
$\tilde{V}_{8R}$  & $+$ & 0 & 0 & 0  &
$+$ & $+$ & $+$ & $\surd$ \\[0.1cm]
$\tilde{V}_{8L}$  & 0 & $+$ & 0 & 0  &
$+$ & $-$ & $-$ & $\surd$ \\[0.1cm]
\hline
& & & & & & & & \\[-0.3cm]
$\tilde{S}_{1R}$  & $0$ & $0$ & $0$ & $-$ &
$+$ & $+$ & $-$ & $\surd$ \\[0.1cm]
$\tilde{S}_{1L}$  & $0$ & $0$ & $-$ & $0$ &
$+$ & $-$ & $+$ & $\surd$ \\[0.1cm]
$\tilde{S}_{8R}$  & $0$ & $0$ & $0$ & $+$ &
$-$ & $-$ & $+$ & $\times$ \\[0.1cm]
$\tilde{S}_{8L}$  & $0$ & $0$ & $+$ & $0$ &
$-$ & $+$ & $-$ & $\times$ \\[0.1cm]
\hline
& & & & & & & & \\[-0.3cm]
$S_2^\alpha$ & $-$ & $0$ & $0$ & $0$ &
$-$ & $-$ & $-$ & $\times$ \\[0.1cm]
$S_{13}^{\alpha\beta}$  & $+$ & $0$ & $0$ & $0$ &
$+$ & $+$ & $+$ & $\surd$ \\[0.1cm]
\hline
& & & & & & & & \\[-0.3cm]
$V_{8R}$ & $\pm$ & $0$ & $0$ & $0$ &
$\pm$ & $\pm$ & $\pm$ & $\surd (+)$ or $\times (-)$ \\[0.1cm]
$V_{8L}$ & $0$ & $\pm$ & $0$ & $0$ &
$\pm$ & $\mp$ & $\mp$ & $\surd (+)$ or $\times (-)$ \\[0.1cm]
$V_{8R}\,,V_{8L}$ & indef. & indef. & indef. & indef. &
indef. & indef. & indef. & indef. \\[0.1cm]
\hline\hline
\end{tabular}
\end{center}
\end{table*}

\begin{figure}
\hspace{-1.5cm}
\includegraphics[width=10cm]{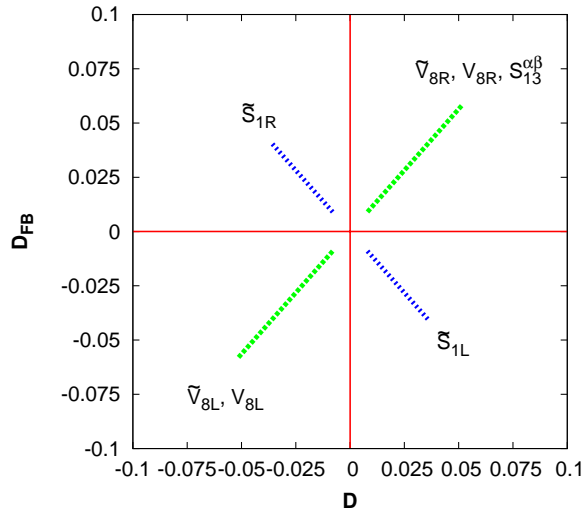}
\caption{
The predictions for $D$ and $D_{\rm FB}$ of the models 
under consideration, being
consistent with the $\sigma_{t\bar{t}}$ and $A_{\rm FB}$ 
measurements at the 1-$\sigma$ level. We assume only one resonance
exists or dominates.}
\label{fig:ddfb_onecoupling}
\end{figure}

4. Let us first consider the FV cases.
Among the FV interactions with vector or scalar bosons, 
$\tilde{V}_{8R,8L}$, $\tilde{S}_{1R,1L}$, and $S_{13}^{\alpha\beta}$ can give 
the correct sign for $(C_1 - C_2)$~\cite{jkln1}.
But one can not discriminate one model from another only 
with the $A_{\rm FB}$ measurement. 
From Table~\ref{tab:newparticles}, we observe that each of the four cases
with $\tilde{V}_{8R}$, $\tilde{V}_{8L}$, $\tilde{S}_{1R}$, and $\tilde{S}_{1L}$ 
gives a different sign combination of $C_1^\prime+C_2^\prime$ and 
$C_1^\prime-C_2^\prime$.
Therefore, a simple sign measurement of $D$ and $D_{\rm FB}$ can 
endow us with the model-discriminating power.
In Fig.~\ref{fig:ddfb_onecoupling}, we show the prediction of each model for
$D$ and $D_{\rm FB}$ varying the model parameters  over the ranges:
\begin{eqnarray}
\tilde{V}_{8R,8L}    &:&
  \frac{1}{N_c}\,\left(\frac{1\,{\rm TeV}}{m_{\tilde{V}_{8R,8L}}}\right)^2\,
|\tilde{g}_{8q}^{R,L} |^2 \simeq 0.56 \pm 0.41 \,, \nonumber \\
\tilde{S}_{1R,1L}    &:&
  \left(\frac{1\,{\rm TeV}}{m_{\tilde{S}_{1R,1L}}}\right)^2\,
 |\tilde{\eta}_{1q}^{R,L} |^2 \simeq 0.41 \pm 0.26 \,, \nonumber \\
{S}_{13}^{\alpha\beta}    &:&
  2\,\left(\frac{1\,{\rm TeV}}{m_{S_6}}\right)^2\, |\eta_6|^2 
\simeq 0.56 \pm 0.41 \ .
\end{eqnarray}
which are consistent with the current measurements
of $\sigma_{t\bar{t}}$ and $A_{\rm FB}$ at the 1-$\sigma$ level 
(see Fig.~\ref{fig:newfig1}).
We observe that $D$ and $D_{\rm FB}$ take the same $(+,+)$
and $(-,-)$ signs for $\tilde{V}_{8R}$ and
$\tilde{V}_{8L}$, respectively, while they take
the different $(+,-)$ and $(-,+)$ signs for
$\tilde{S}_{1L}$ and $\tilde{S}_{1R}$, respectively.
The color-sextet scalar $S_{13}^{\alpha\beta}$ gives
the same $(+,+)$ sign as the $\tilde{V}_{8R}$ case.

Unlike the FV cases, the FC color-octet 
vectors can always accommodate the positive sign of $(C_1 - C_2)$.
For the case of $V_{8R}$ ($V_{8L}$), the couplings $g^R_{8q}$ ($g^L_{8q}$) 
and $g^R_{8t}$ ($g^L_{8t}$) must have different signs to 
accommodate the positive $A_{\rm FB}$. In Fig.~\ref{fig:ddfb_onecoupling}, 
we also show the predictions of the model with $V_{8R}$ or $V_{8L}$ vector
for $D$ and $D_{\rm FB}$.
%

\begin{figure}
\hspace{-1.5cm}
\includegraphics[width=10cm]{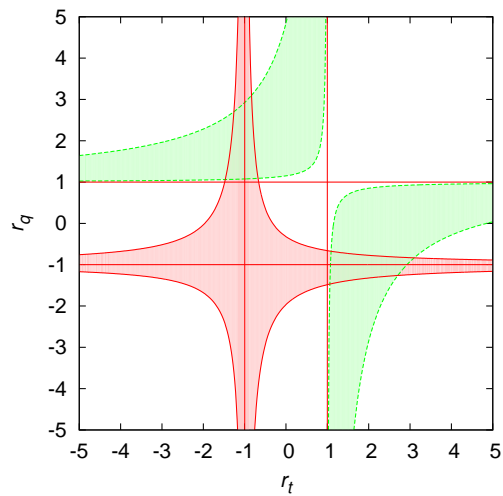}
\caption{
The region in $(r_t , r_q)$ plane that is
consistent with the Tevatron
$\sigma_{t\bar{t}}$ (red) and $A_{\rm FB}$ (green)
measurements at the 1-$\sigma$ level.
The general flavor-conserving case is considered taking
$g^L_{8q}g^L_{8t}\,(1\,{\rm TeV}/m_{V8})^2=+1$.
}
\label{fig:rqrt_plusone}
\end{figure}
\begin{figure}
\hspace{-1.5cm}
\includegraphics[width=10cm]{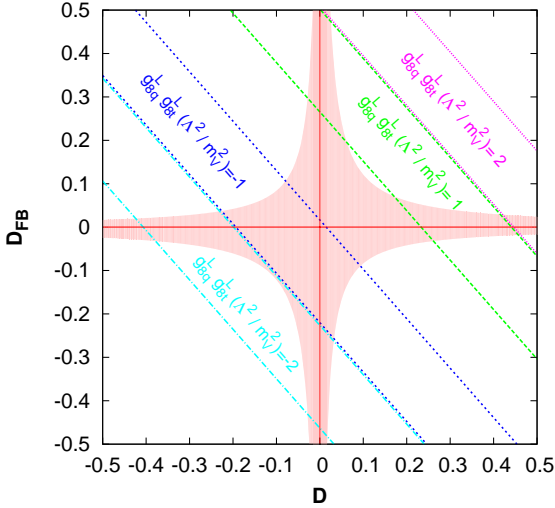}
\caption{
The predictions for $D$ and $D_{\rm FB}$, being
consistent with the $\sigma_{t\bar{t}}$ and $A_{\rm FB}$
measurements at the 1-$\sigma$ level, for several values
of $g^L_{8q}g^L_{8t}\,(1\,{\rm TeV}/m_{V8})^2=$
$+2$ (magenta), $+1$ (green), $-1$ (blue), and $-1$ (sky blue),
from the upper-right corner to the lower-left one.
The general model with
flavor-conserving color-octet
$V_{8R}$ and $V_{8L}$ vectors is considered. }
\label{fig:ddfb_v8}
\end{figure}
%
5. Up to now,  
we have only one type of couplings by assuming that 
only one resonance contributes to the $t\bar{t}$ 
production at the Tevatron.
However, the flavor-conserving color-octet
$V_{8R}$ and $V_{8L}$ vectors can coexist in general, 
and then the situation could be more complicated. 
In such a general case, all the four couplings  
$C^{RR}$, $C^{LL}$,$C^{LR}$, and $C^{RR}$ could be nonzero,  
in contrast to the previous one-coupling cases. 
In this case, the sum and differences of the couplings can be 
reparametrized as
\begin{eqnarray}
(C_1+C_2)/\Lambda^2 &=& 
-g^{L}_{8q}g^{L}_{8t}(r_q+1)(r_t+1)/m_{V8}^2
\nonumber \\[0.1cm]
(C_1-C_2)/\Lambda^2 &=&
-g^{L}_{8q}g^{L}_{8t}(r_q-1)(r_t-1)/m_{V8}^2
\nonumber \\[0.1cm]
(C_1^\prime+C_2^\prime)/\Lambda^2 &=&
-g^{L}_{8q}g^{L}_{8t}(r_q+1)(r_t-1)/m_{V8}^2
\nonumber \\[0.1cm]
(C_1^\prime-C_2^\prime)/\Lambda^2 &=&
-g^{L}_{8q}g^{L}_{8t}(r_q-1)(r_t+1)/m_{V8}^2
\end{eqnarray}
with $r_q \equiv g^{R}_{8q}/g^{L}_{8q}$ and
$r_t \equiv g^{R}_{8t}/g^{L}_{8t}$. 
Any deviation of $r_q$ ($r_t$) from $1$ 
characterizes $P$ violation in the light (top) quark sector.
In Fig.~\ref{fig:rqrt_plusone}, we show 
the 1-$\sigma$ region in $(r_t , r_q)$ plane 
taking $g^L_{8q}g^L_{8t}\,(1\,{\rm TeV}/m_{V8})^2=+1$.
We observe the consistent region lies along the line
$r_t=-1$ ($r_q=-1$) with $1<r_q\lsim 3$ ($1<r_t\lsim 3$).
When $g^L_{8q}g^L_{8t}\,(1\,{\rm TeV}/m_{V8})^2=-1$,
one may have obtain similar results, except that the green region 
consistent with $A_{\rm FB}$ would be reflected 
with respect to the $r_q=1$ line. 
In Fig.~\ref{fig:ddfb_v8}, we show
the predictions of the general model with $V_{8R}$ and $V_{8L}$
for $D$ and $D_{\rm FB}$ taking
$g^L_{8q}g^L_{8t}\,(1\,{\rm TeV}/m_{V8})^2=\pm 1\,,\pm 2$.
Note that the experimental measurements on the
$\sigma_{t\bar{t}}$ and $A_{\rm FB}$ 
constrains the product of $D$ and $D_{\rm FB}$ independently of
$g^L_{8q}g^L_{8t}\,(\Lambda/m_{V8})^2$. 
This can be easily understood by observing the relation
$(C_1+C_2)(C_1-C_2) = 
(C_1^\prime+C_2^\prime) (C_1^\prime-C_2^\prime)$ which leads to
\begin{equation}
\Delta\sigma_{t\bar{t}}\,\Delta A_{\rm FB}
\propto D\,D_{\rm FB}\,.
\end{equation}
Let us note that 
$\Delta\sigma_{t\bar{t}} \propto (C_1+C_2)$,
$\Delta A_{\rm FB} \propto (C_1+C_2)$,
$D \propto (C_1^\prime+C_2^\prime)$, and
$D_{\rm FB} \propto (C_1^\prime-C_2^\prime)$.
Furthermore, we observe
\begin{eqnarray}
&&\hspace{-0.5cm}
g^L_{8q}g^L_{8t}\,\left(\frac{\Lambda}{m_{V8}}\right)^2
\\[0.1cm]
&=&
\frac{\left[(C_1^\prime+C_2^\prime)-(C_1+C_2)\right]\,
\left[(C_1^\prime+C_2^\prime)-(C_1-C_2)\right]}
{4(C_1^\prime+C_2^\prime)}
\nonumber \\[0.1cm]
&=&\frac{1}{4}\left[
(C_1^\prime+C_2^\prime) - (C_1+C_2) -(C_1-C_2) 
+(C_1^\prime-C_2^\prime)\right]\,,\nonumber
\end{eqnarray}
where, for the last term, the relation $(C_1+C_2)(C_1-C_2) =
(C_1^\prime+C_2^\prime) (C_1^\prime-C_2^\prime)$ is used.
This explains the linear dependence of
$D$ and $D_{\rm FB}$  on 
$g^L_{8q}g^L_{8t}\,\left({\Lambda}/{m_{V8}}\right)^2$
with some finite range coming from the 
current 1-$\sigma$ experimental errors
on $\sigma_{t\bar{t}}$ and $A_{\rm FB}$,
as shown in  Fig.~\ref{fig:ddfb_v8}.
We see that one of
$|D|$ and $|D_{\rm FB}|$ could be as large as $\sim 1$
when the other one is very small,
while both of them could be $\sim 0.1$ simultaneously.
%


6. In this letter,  we extended the model independent study of 
of $t\bar{t}$ productions at the Tevatron 
using dimension-6 contact interactions relevant to 
$q\bar{q}\rightarrow t \bar{t}$, mainly concentrating on the 
longitudinal (anti)top polarization of $P_L$ and $\bar{P}_L$ 
in the helicity frame.  As emphasized in Ref.~\cite{jkln1}, new physics
affecting the Tevatron $A_{\rm FB}$  necessarily breaks parity 
unlike QCD. Then the $P$-odd top-quark longitudinal polarization 
observables can be nonzero, in sharp contrast to the case of pure QCD. 
Therefore, nonvanishing longitudinal polarization observables  
will be another important aspect  of $P-$violating new physics 
relevant to $q\bar{q} \rightarrow t\bar{t}$. 
Most importantly, the longitudinal polarization of (anti)top quark can give 
another important clue for the chiral structure of new physics, 
which is completely independent of 
$\sigma_{t\bar{t}}$ or $A_{\rm FB}$.

Using the conditions for the couplings of four-quark operators 
that could generate the FB asymmetry observed at the Tevatron 
(with the updated data on $A_{\rm FB}$) \cite{jkln1}, 
we studied the possible ranges of longitudinal (anti)top polarization, 
and their correlations with $\sigma_{t\bar{t}}$ and $A_{\rm FB}$. 
Then we considered the $s-$, $t-$ and $u-$channel exchanges of 
spin-0 and spin-1 particles whose color quantum number is 
either singlet, octet, triplet or sextet.  Our results in Table~I  
encode the predictions for the $P$-odd observables 
corresponding to the polarization difference 
$(P_L - \bar{P}_L)$ in various new physics scenarios 
in a compact and an effective way, when those new particles
are too heavy to be produced at the Tevatron but still affect $A_{\rm FB}$.  
If these new particles could be produced directly at the Tevatron or 
at the LHC,  we cannot use the effective lagrangian any more. 
We have to study specific models case by case including the new particles
explicitly, and anticipate rich phenomenology at colliders as well as 
at low energy. Detailed study of these issues lies beyond the scope 
of this letter, and will be discussed in the future publications \cite{progress}.

{\it Note Added:} 
While we were finishing this paper, we received 
three preprints \cite{Godbole:2010kr,Degrande:2010kt,Cao:2010nw}  
which also consider the observables 
related with the (anti)top polarization.  
In our work, we note that parity violation is crucial 
for new physics to make nonzero contributions to $A_{\rm FB}$, 
and the longitudinal polarization of (anti)top quark can give 
another important clue for the chiral structure of new physics. 


\begin{acknowledgements}
We are grateful to S. Choi, D.H. Kim, Hyunsoo Kim, S.B. Kim, 
H.S. Lee and I. Yu for useful communications, and 
S.-h. Nam for collaboration in the early stage of this work. 
We thank Dan Amidei for 
the prospects for measuring  the
top-quark helicity cross sections at the Tevatron.
Part of works by PK were done at Aspen Center for Physics.   
\end{acknowledgements}



%

\end{document}